# Graph Sampling Approach for Reducing Computational Complexity of Large-Scale Social Network

**Andry Alamsyah[1,2], Yahya Peranginangin[1], Intan Muchtadi-Alamsyah[3], Budi Rahardjo[2] and Kuspriyanto[2]**

[1] School of Economics and Business, Telkom University, Indonesia
[2] School of Electrical Engineering, Bandung Institute of Technology, Indonesia
[3] Faculty of Mathematics and Natural Science Bandung Institute of Technology Indonesia

## Abstract

The online social network services provide platform for human social interactions. Nowadays, many kinds of online interactions generate large-scale social network data. Network analysis helps to mine knowledge and pattern from relationship between actors inside the network. This approach is important to support predictions and decision-making process in many real-world applications. The social network analysis methodology, which borrow approach from graph theory provides several metrics that enabled us to measure specific properties of the networks. Some of metrics calculations were built with no scalability in minds, thus it is computationally expensive. In this paper, we propose graph sampling approach to reduce social network size, thus reducing computation operations. The performance comparison between natural graph sampling strategies using edge random sampling, node random sampling and random walks are presented on each selected graph property. We found that the performance of graph sampling strategies depends on graph properties measured.

# 1 Introduction

Our daily online interactions have contributed to the rise of data production. This lead to massive data available, which in turn can offer opportunity for us to model the phenomenon and make predictions that support decision making process in many real world applications. The legacy methodology in social sciences are based on directly working with the population and few take the route to gain insight using massive data available in online.

We use the *Social Network Analysis* (SNA*)* [1] [2] methodology to analyze the network data. SNA foundation are based on graph theory [3]. SNA provides many metrics that support network topology measurement but some of them for example *Betweenness Centrality* were built for small network. As the network becomes bigger, it increases the computational complexity of finding the shortest path between all pairs in network. *Betweenness Centrality* have computational complexity reaching $O(|N|^3)$ [4], where $|N|$ is the number of nodes in the network.

In our previous attempt to reduce graph size and make representative sample, we propose *Graph Summary* [5] based on *Minimum Description Length* [6] principle. Our effort can reduce the graph size by 50%, but the time needed to construct graph summary is high, because of merging nodes operation complexity. For that reason, we try another alternative with more simple approach such as using graph sampling methodology.

Graph sampling [7] is a technique to pick a subset of nodes and/or edges from original graph. Although being smaller in size, sampled graph may be similar to original graph in some way (see. Figure 1). In this paper, we are interested on how properties behave after graph sampling implementation. We generate different size of graph sample in order to see the properties evolution in different size of network. We investigate which properties are preserved in certain sampling technique and which properties are not. If a property in sampled graph is preserved, then we can estimate the property value in original network.

We define original graph $G(N_G, E_G)$ where $N_G$ is set of nodes and $E_G$ is set of edges. The graph sample $S(N_S, E_S)$ is subset of $G(N_G, E_G)$, where $N_S \quad N_G$ and $E_S \quad E_G$. We denote $|N|$ as the number of nodes and $|E|$ as the number of edges.

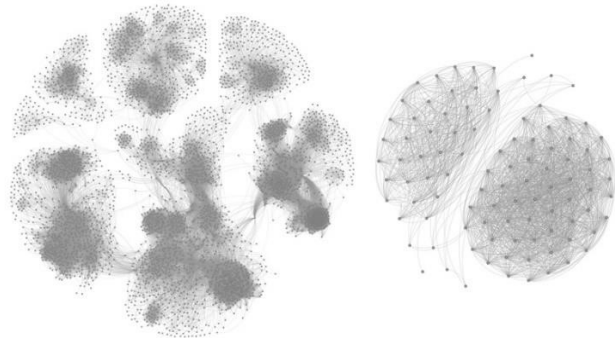

Figure 1. (a) left: Original network with 4039 nodes and 88234 edges. (b). right: Sampling of original network using *RW* strategy with 100 nodes and 2013 edges

## 2 Social Network Properties

There are many social network metrics available to describe certain social network properties. We use the following properties [2] as comparison in each graph sample size.

1.  *Average Degree* is the number of edges *E* compared to number of nodes *N*. We denote *Average Degree = |E|/|N|*
2.  *Density* is the number of edges *E* in the network compared to maximum number of edges between nodes in the network. We denote *Density = |E|/(|N|\*(|N|-1))*
3.  *Modularity* measures fraction of the edges that fall within the given groups minus the expected of such fraction is edges were distributed at random. The bigger *M* value means the boundary between groups in the network are more distinct.
4.  *Average Clustering Coefficient* measures the total degree to which nodes are tend to cluster together in network compared to total number of nodes.
5.  *Diameter* measures the longest of shortest path between any pair of nodes in network.
6.  *Average Path Length* calculated by finding the shortest path between all pairs of nodes, adding them up and then dividing by the total number of pairs.
7.  *Connected Component* shows how many network components in which any two nodes are connected to each other by paths.

## 3 Graph Sampling Strategies

We propose three strategies of graph sampling by using *Edge Random Sampling (ERS)*, *Node Random Sampling (NRS)* and *Random Walks Sampling (RW)* methods. We explain each of the strategy and algorithm construction as follows:

### 3.1 Edge Random Sampling

The *Edge Random Sampling (ERS)* strategy is the simplest strategy of three. We pick random edges uniformly from the network. The selected edges are used to construct new smaller network. The *ERS* algorithm is as follows:

```
1:    input network G(N_G, E_G)
2:    choose set of random edges E_S from G
3:    construct network sample S(N_S, E_S), where E_S are the chosen edges
      from E_G and N_S are selected nodes connecting the E_S
```

### 3.2 Node Random Sampling

The *Node Random Sampling (NRS)* strategy works as follows. We pick random nodes uniformly from the network, then we construct new smaller network from the chosen nodes. The *NRS* algorithm is as follows:

```
1:    input network G(N_G, E_G)
2:    choose set of random nodes N_C from G
3:    create permutations list P = P(N_G, 2)
4:    create intersection list I = P & G , where I is the edge list from the
      connections between selected N_G from G
5:    construct network sample S(N_S, E_S) from list I
```

### 3.3 Random Walks Sampling

The *Random Walks (RW)* strategy are based on random walks idea, where given a network and a starting node, then we select neighbor of this node at random and move to it. The random sequence selected, which consists of nodes and edges are kept as the result of *RW* strategy.

We run each *RW* process for certain number of iterations which we choose carefully depends what we see as the sufficient sample of network size. The *x* most visited nodes from the iterations become our candidate for network sample. From this point, the process similar with NRS process. The difference *RW* process and *NRS* process is nodes are not chosen by uniformly random sampling, but by random walks mechanism. The *RW* algorithm is as follows:

```
1: input network G(N_G, E_G)
2: while r < number of run
3: while i < number of iterations: We randomly choose N_G(i+1) where N_G(i+1) is
neighbor of N_G(i)
4: list L_r = (N_G(i), N_G(i+1),N_G(i+2),…,N_G(number of iterations))
5: list L = L_1 + L_2 + …. + L_number of run
6: list M = descending sorted list L by number times a node has been selected in RW
strategy
7: list M_x = x most visited node
8: create permutations list P = P(M_x, 2)
9: create intersection list I = P & G , where I is the edge list from the
connections between selected N_G from G
10: construct network sample S(N_S, E_S) from list I
```

## 4 Experiments

We use *ego-Facebook* dataset from *Stanford Network Analysis Project* (SNAP) repositories for our experiment to construct graph sample, using three strategies in section 3. This dataset network consists of *4039* nodes and *88234* edges. The dataset network properties are *Density 0.011, Average Degree 43,691, Modularity 0.835, Average Clustering Coefficient 0.617, Diameter 8, Average Path Length 3.693 and Connected Component 1*

The *ERS* algorithm experiment is run by pick a set of random edges which each round chosen by stepping up *10000* edges from *10000* edges to *70000* edges. The random sampling nature of this algorithm produces fluctuate values. To overcome discrepancy between each result, on each round, we run the algorithm *10* times. The average network properties value from all run become our final result of the *ERS* strategy.

The *NRS* algorithm experiment is run by pick some randomly uniform nodes, start from *100* nodes, and then continue to *500, 1000, 1500, 2000, 2500, 3000, 3500* nodes. This set of nodes $N_C$ become candidate of our graph sample. To check whether there is a connection between a pair of selected nodes, we construct permutations mechanism $P(N_C, 2)$ , where we list all possible permutations between pairs of nodes. On the last step, we check whether there is actual connection between a pair of nodes from the list of possible connection between nodes resulted from permutations mechanism $P$ by using intersection list $I$ between list of $N_G$ and list of $N_C$. The list $I$ converted to edge list and become network sample $S(N_S, E_S)$.

The *RW* algorithm experiment is run with the scenario as follows. At one random walks process, we set the number of iterations $i$ as the numbers of how many nodes collected on each random walks process. We choose the value $i = 10000$, as we consider this number is suitable enough to get representative number of nodes. We run *10* random walks process. On each random walks process, we pick $x$ most visited nodes, throughout the experimentation we change variable $x$ to *100, 500, 1000, 1500, 2000, 2500,* and *3000* nodes respectively. The sum of all 10 random walks process are collected, and we pick the top *100, 500, 1000, 1500, 2000, 2500* and *3000* most visited nodes respectively. The network from the top most visited nodes becomes our graph sample $S(N_S, E_S)$.

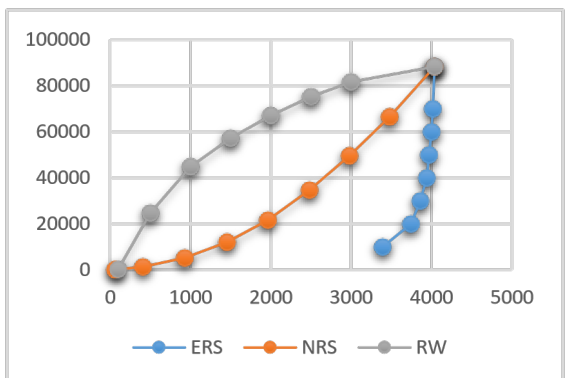

(a). Graph Sample Size (nodes vs edges)

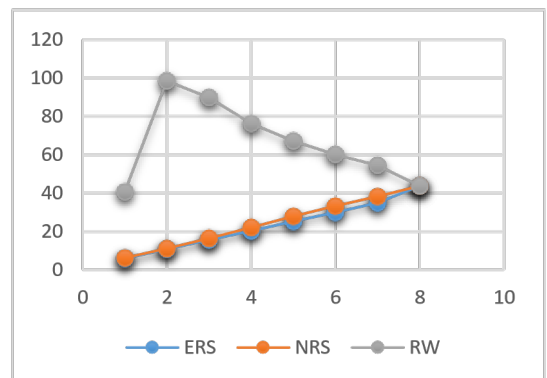

(b). Average Degree (size vs value)

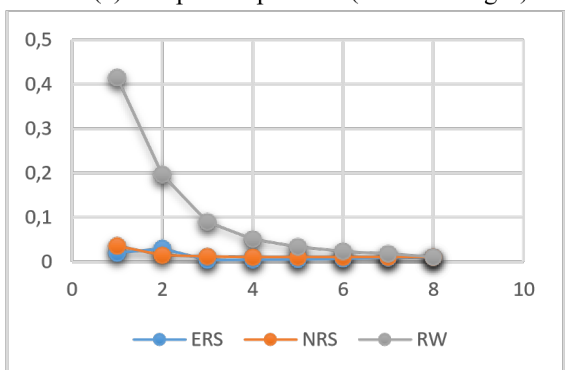

(c). Density (size vs value)

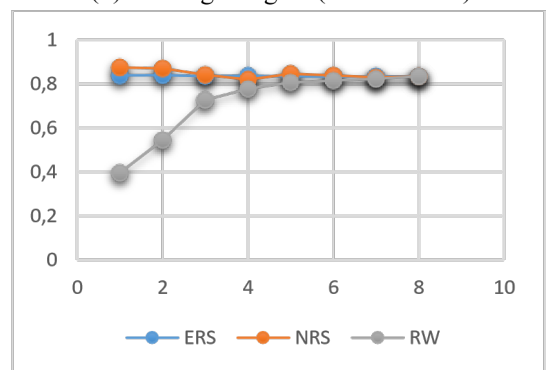

(d). Modularity (size vs value)

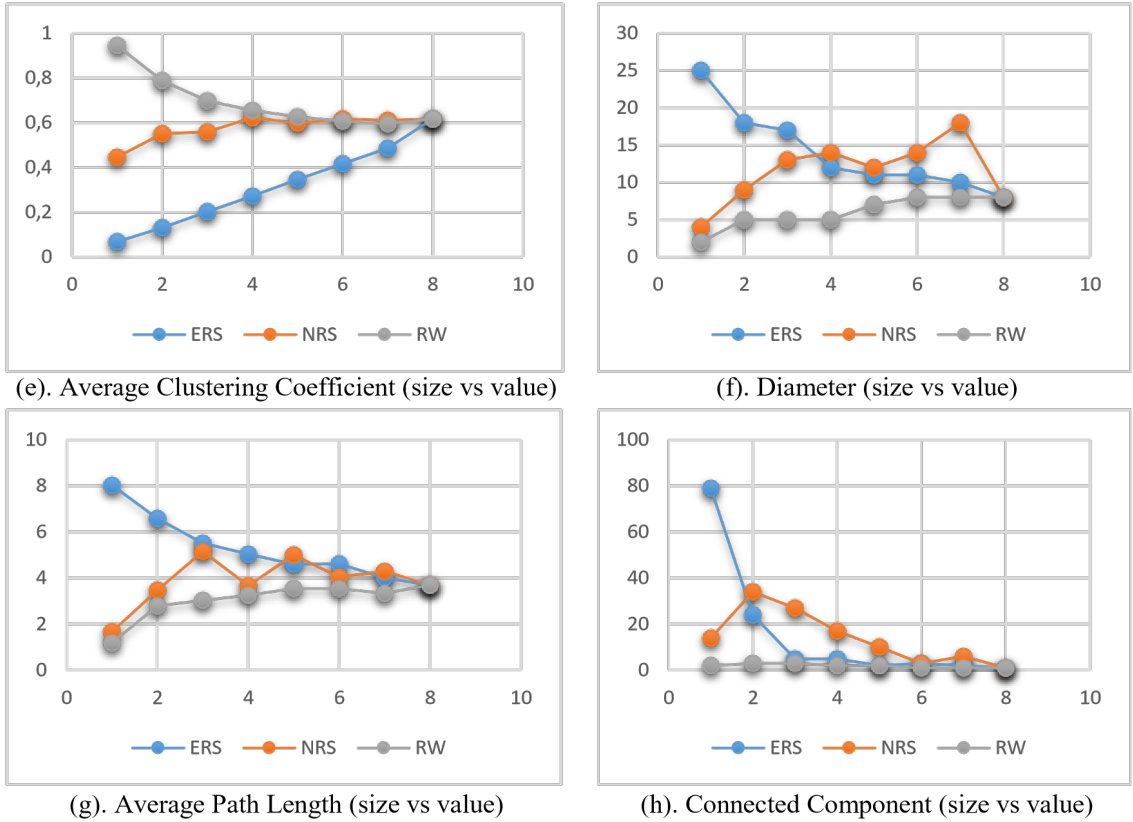

(e). Average Clustering Coefficient (size vs value)

(f). Diameter (size vs value)

(g). Average Path Length (size vs value)

(h). Connected Component (size vs value)

Figure 2. Experiment result chart (a). Graph Sample Size, (b). Average Degree, (c). Density, (d). Modularity, (e). Average Clustering Coefficient, (f). Diameter, (g). Average Path Length, (h). Connected Component

## 5 Discussions and Conclusions

Figure 2 show the performance chart of three sampling methods. In *Graph Sample Size*, *RW* create sample graph with smaller number of nodes but with much bigger number of edges compared to the other two, this signify that sample from *RW* are faster and more effective in constructing representative graph sample. A method performs better in a graph property if, 1). it converges to the value of the original graph property with predicted result, both linearly or exponentially and 2). the gap between predicted value and the original value is small. With this regards, *RW* methods are performs better in properties *Average Clustering Coefficient, Diameter, Average Path Length, Connected Component*. Both *NRS* and *ERS* are performs better in properties *Average Degree*, *Modularity* and *Density*.

To reduce the computational complexity of *Betweenness Centrality* metrics as the sample of our original problems. We pick a property that related to our problem, in this case is finding shortest path between all pairs. The problem is related to *Average Path Length* and *Diameter* properties; thus we choose *RW* methods.

Our conclusions are 1). We can use graph sampling based on three strategies to reduce graph size while still retain some properties or retain some information that we use as estimator of original graph properties. 2). Different graph properties need different graph sampling methodology. 3). For the future works, we need to compare the computations complexity of each sampling methods in order to get the conclusive results regarding each sampling methods.

## References

[1]  J. Scott, *Social Network Analysis: a Handbook*, Sage Publications, 2000.

[2]  M.E.J. Newman, *Network: An Introduction*, Oxford University Press, 2010. http://dx.doi.org/10.1093/acprof:oso/9780199206650.001.0001

[3]  R. Diestel, *Graph Theory: Electronic,* Edition 2005, Springer-Verlag Heidelberg, New York, 1997, 2000, 2005.

[4]  U. Brandes, A Faster Algorithm for Betweenness Centrality, *Journal of Mathematical Sociology*, **25** (2000), no. 2, 163-177. http://dx.doi.org/10.1080/0022250x.2001.9990249

[5]  A. Alamsyah, Y. Peranginangin, B. Rahardjo, I. Muchtadi-Alamsyah, Kuspriyanto Kuspriyanto, Reducing Computational Complexity of Network Analysis using Graph Compression Method for Brand Awareness Effort, *The 3rd International Conference on Computational Science and Technology*, (2015), 1-6. http://dx.doi.org/10.2991/iccst-15.2015.26

[6]  P.D. Grunwald, *The Minimum Description Length Principle,* The MIT Press, 2007.

[7]  P. Hu, W.C. Lau, *A Survey and Taxonomy of Graph Sampling*, arXiv:1308.5865[cs.SI], 2013.